**Title**:
Twirling of actin by myosins II and V observed via polarized TIRF in a modified gliding assay

**Running Title:**
Twirling of actin by myosins II and V

**Authors**:
John F. Beausang*
Harry W. Schroeder III[#]
Philip C. Nelson*
Yale E. Goldman[§¶]

**Affiliations**
* Department of Physics and Astronomy
# Biochemistry and Molecular Biophysics
[§] Department of Physiology
[¶] Pennsylvania Muscle Institute

University of Pennsylvania
Philadelphia, PA 19104

## ABSTRACT

The force generated between actin and myosin acts predominantly along the direction of the actin filament, resulting in relative sliding of the thick and thin filaments in muscle or transport of myosin cargos along actin tracks. Previous studies have also detected lateral forces or torques that are generated between actin and myosin, but the origin and biological role of these sideways forces is not known. Here we adapt an actin gliding filament assay in order to measure the rotation of an actin filament about its axis ("twirling") as it is translocated by myosin. We quantify the rotation by determining the orientation of sparsely incorporated rhodamine-labeled actin monomers, using polarized total internal reflection (polTIRF) microscopy. In order to determine the handedness of the filament rotation, linear incident polarizations in between the standard $s$- and $p$-polarizations were generated, decreasing the ambiguity of our probe orientation measurement four-fold. We found that whole myosin II and myosin V both twirl actin with a relatively long (~ μm), left-handed pitch that is insensitive to myosin concentration, filament length and filament velocity.

## INTRODUCTION

The swinging lever arm model (1, 2) explains force production and movement between actin and myosin II in muscle contraction (3) and also applies to non-muscle myosins (4). Separate crystal structures of myosin and actin docked into cyro-EM maps of actomyosin (5) indicate that the lever arm swing is nearly parallel to the axis of the actin filament, thereby efficiently converting the free energy released from ATP hydrolysis into motion along the filament. Even small torque components around the filament axis, however, may have biological roles in muscle contraction ((6), and references therein) or regulation (7). For processive non-muscle motors, a torque may be desirable for navigating cargo around obstacles present in the crowded environment of the cell (8-10).

Several studies have suggested off-axis components to the relative motion between actin and myosin. For example, decreased lateral spacing of the filaments of frog skeletal muscle in rigor compared with relaxation was attributed to radial forces between the thick and thin filaments (11). In a modified gliding assay, actin filaments that were selectively immobilized onto the slide at their pointed ends formed superhelices that suggested a right-handed component of torque generated by myosin II (12). In standard gliding assays where the filaments are free to translocate, a torque component of the cross-bridge force could result in rotation of the filament about its longitudinal axis, a motion we term "twirling." Actin filaments with marker beads attached at their ends showed no twirling on HMM-coated slides (13) while gliding, but when the filaments were marked sparsely with fluorescently labeled actin monomers, simultaneous twirling and gliding of the filaments was observed by polarized fluorescence microscopy (14, 15). Symmetries in the fluorescence polarization technique prevented both of those studies from determining the handedness of the twirling motion.

A separate way of gauging sideways motions and possible components of torque between actin and processive myosins is to suspend the actin filament above the surface of the microscope slide and record the path of a bead being transported by myosin along the suspended filament (16, 17). Off-axis force causes the bead to travel in a helical path. In the suspended filament assay, myosins V (16) and VI (17) exhibited left-handed and right-handed helical paths, respectively.

Polarized epifluorescence microscopy (14) and polarized total internal reflection fluorescence (polTIRF) microscopy (15, 18, 19) typically use incident and/or detected light polarized along the $x$, $y$ and $z$ axes of the microscope. In such configurations, rotation of the fluorophore, but not its handedness, can be observed because orientations of the fluorophore reflected across any of the Cartesian planes give the same fluorescence intensities and thus are not distinguishable. Intermediate excitation or emission polarizations that break these symmetries enable the handedness to be recovered (20).

In the present work, we added extra linear polarizations to the previously reported polTIRF technique (18, 19, 21, 22) that are intermediate between those aligned along the Cartesian directions. As a result, there is a four-fold increase in the range of unambiguously detected probe orientations. The orientation is then estimated within a hemisphere, the remaining two-fold ambiguity being a property of dipolar absorption and emission of light.

We report twirling and its handedness from gliding actin filaments that are translocated by whole myosin II and myosin V. For myosin II, the twirling motion is nearly always left-handed with an average pitch, i.e. the distance that the filament translocates during one complete rotation, of approximately 0.5 μm that is not strongly influenced by myosin concentration,

MgATP concentration or filament length in the range studied here. Twirling by native myosin V was also found to be left-handed with an average pitch of 1.7 µm. These values for the twirling pitch are much longer and opposite in handedness to the intrinsic pitch of the actin filament. Several mechanical effects are discussed that could give rise to filament twirling. The results have been presented in abstract form (23).

## METHODS

### Biological Samples

Proteins, buffers, and slides were prepared as in (19) and (22) with minor modifications. Briefly, whole myosin II was purified from rabbit fast skeletal muscle and stored in 300 mM KCl, 5 mM Hepes, pH 7.0, 5 mM $NaN_3$ and 50% glycerol at -20°C (24). Native myosin V was obtained from brains of 1-day old chicks (25). Actin was prepared from rabbit muscle in G-buffer (2 mM Tris, pH 8.0, 0.2 mM $CaCl_2$, 0.2 mM ATP, 0.5 mM DTT), frozen in liquid $N_2$, and stored at -80°C (26). G-actin monomers were labeled (27) at $Cys^{374}$ with 5′-iodoacetamido-tetramethylrhodamine (a gift from J. E. T. Corrie, National Institute for Medical Research, Mill Hill, London). Filaments, sparsely-labeled (~0.05%) with rhodamine-actin, were polymerized by mixing labeled and unlabeled monomers, at 1 µM total actin concentration, in F-buffer (50 mM KCl, 2 mM $MgCl_2$, 1 mM EGTA, 10 mM HEPES, pH 7.4) and stabilized with 1.1 µM Alexa 647 labeled phalloidin (Invitrogen A22287).

In order to remove "dead" myosin heads that do not release actin upon binding ATP, a 12 mg/ml myosin stock was diluted in high salt buffer (500 mM KCl, 10 mM Hepes, pH 7.0, 5 mM $MgCl_2$) and combined with a molar excess of actin filaments, 2.5 mM ATP and 5 mM DTT and then centrifuged at 200,000 × g for 30 min on the day of each experiment. The concentration of protein retained in the supernatant was then determined by Bradford assay.

A ~20 µl flow cell was created using a clean quartz microscope slide, glass cover slip, and two pieces of double-sided tape. For myosin II, a 1 mg/ml solution of poly-L-lysine was flowed into the cell and incubated for 1 min. Excess polylysine was rinsed out with 20 µl of high salt buffer, and then the myosin, diluted from stock in high salt buffer, was flowed into the cell and incubated for 2 min. The following solutions were made in wash buffer (WB: 25 mM KCl, 20 mM Hepes pH 7.4, 5 mM $MgCl_2$) and flowed through the cell. 2 × 0.5 mg/ml BSA, 2 × 5 µM unlabeled sheared actin filaments to block any remaining dead myosin heads, 2 mM ATP to dissociate actin from the active heads, and 2 × WB to wash out the free actin and excess ATP. Finally, 5 nM 0.05% rhodamine-labeled Alexa 647 phalloidin-stabilized actin filaments and motility buffer with 5-20 µM ATP, 50 mM DTT, 5 mM phosphocreatine and 0.4 mg/ml creatinephosphokinase were flowed into the cell. All experiments were performed at 22-23 °C.

The concentration of myosin II flowed into the cell was equal to 0.03, 0.1, 1, or 3 mg/ml. MgATP concentration was varied between 5 and 20 µM, which resulted in filament velocities of 0.1-0.5 µm/s. For low myosin concentrations (≤ 0.1 mg/ml), 0.1% methylcellulose was added to the motility solution to increase filament run length. Twirling of filaments with variable lengths in the range 1-50 µm was determined at concentrations of 0.03 and 0.1 mg/ml.

Myosin V twirling assays were similar to those with myosin II except for the elimination of the centrifugation step to remove dead heads, the inclusion of 0.1 mg/ml calmodulin (28), and higher MgATP concentrations (typically 0.5 -1 mM). For myosin V, experiments were performed at 0.26 mg/ml loading myosin concentrations.

## Experimental Setup

The polarized total internal reflection fluorescence (polTIRF) setup is described elsewhere (19) and (22). Here we provide a brief explanation and describe modifications that enable linear polarizations of the incident laser illumination that are polarized in between the *s*- and *p*-polarization directions, see Fig. 1.

Two alternating beams from a 532 nm Nd:YAG laser are focused onto a prism such that the incident angle (with respect to the *z* axis, normal to the quartz slide/water interface) is ~ 68°. Fluorescent emission is collected by a 100× 1.2 NA water immersion lens, passed through a long pass blocking filter and either imaged onto an intensified CCD camera or passed through a polarizing beam-splitter onto two avalanche photodiodes (APDs). Using a Pockels cell (PC0 in Fig. 1) and a polarizing beam splitting cube (PBS0), the incident illumination is cycled between two paths that are aligned predominantly along the *x* and *y* directions and intersect at the sample. Four linear polarizations in each beam path (termed *s*, *p*, *R*, and *L*) are obtained by applying different voltages to additional Pockels cell (PC1 and PC2, New Focus Inc) and then passing the beam through a Berek variable compensator (BC1, BC2) in each path. With 45° between the crystal axis of the PC and the linearly polarized incident light, the PCs produce vertical, horizontal or elliptically polarized light depending on the input voltage. The Berek compensator acts essentially as a ¼-wave plate to convert the elliptically polarized light to beams linearly polarized at any arbitrary azimuth. The bending mirrors used to project the excitation onto the sample re-introduce phase shifts and resulting ellipticity into the intermediately polarized (*L* and *R*) beams. The Berek compensator is adjusted in angle and retardation slightly away from the ¼-wave setting to correct for these extra phase shifts. The *L* and *R* linearly polarized beams are obtained by adjusting the PC voltages and the Berek compensators until the beams are linearly polarized at angles intermediate (approximately ±45°) between *s* and *p* at the entry to the coupling prism (CP). Polarizations are verified with a crossed linear polarizer, which extinguishes the beam immediately prior to the final focusing lens (L1).

Each polarization alternately illuminates the sample for 10 ms in the sequence: *s1, p1, p2, s2, R1, L1, L2,* and *R2*. The fluorescence emission from a selected rhodamine fluorophore is directed through a polarizing beam splitter (PBS1) that separates its *x* and *y* components onto two APDs (APDx and APDy) for a total of 16 measured intensities ($_{s1}I_x$, $_{s1}I_y$, $_{p1}I_x$, $_{p1}I_y$, etc…), see Fig. 2A. For imaging of the field, a mirror (RM) is removed so that the fluorescence is projected onto the intensified CCD camera.

The evanescent field generated by the incident *p* polarized light has a slight (~5%) ellipticity in the *x-z* and *y-z* planes for beams 1 and 2, respectively, due to the component of incident radiation parallel to the reflecting surface. In addition to this *x-z* ellipticity under *p*-polarized illumination, the *L*- and *R*- polarizations in beam 1 have additional ellipticity in the *y-z* plane due to the different phase shifts in the evanescent wave from the *s* and *p* components ($\delta_s$ and $\delta_p$, see Appendix A). A similar phase shift holds for elliptical polarizations from *R2* and *L2*. As done before with ellipticity induced by *p*-polarization (19, 22), the ellipticities of *L* and *R* excitation are included in the analytical equations used to determine the probe orientation.

## Single Molecule Position and Average Filament Velocity

After mounting the sample slide and adding motility buffer, a 30 s movie of a field of ~20-40 single rhodamine fluorophores was recorded by the CCD camera to determine the average filament speed. MgATP concentration was limited to 5 - 20 μM so that the filaments moved slowly enough (0.1-0.5 μm/s) to enable 2-7 s recording of the fluorophore as it passed across the 1.8 µm diameter projected spot size of the APDs before photobleaching. A second movie was recorded after a series of polarization measurements to verify consistent speeds at the beginning and end of the experiment. Occasionally, the first movie was omitted in order to obtain measurements of long filaments, which become shorter due to shearing at high concentrations of myosin. The $(x, y)$ coordinates of the filament were tracked automatically by fitting a 2D Gaussian intensity distribution to the image of the fluorophore for each frame. The $x$-$y$ path of the fluorophore was smoothed using a 5-point Savitsky-Golay filter (29), and the filament velocity was determined from the path length of 15-30 measurements. Average filament velocity for each slide was obtained from 7-10 filaments.

## Single Molecule Orientation and Filament Twirling

Prior to recording each polarization trace, two images of the field of candidate fluorophores were recorded with the CCD camera, superimposed and displayed on a monitor. These images used to calculate a filament-specific velocity for each fluorophore, which could be used as an alternative to the average velocity mentioned above. A molecule was then selected for polarization analysis, and centered above the objective by a computer-controlled piezoelectric stage. The collected fluorescent emission was directed away from the CCD and onto the photodiodes by replacing the removable mirror (RM in Fig. 1). Thus, during polarization recording, spatial information from the fluorophores was not available.

After a moving rhodamine molecule was selected and centered over the APDs, 125 cycles of 16 polarized fluorescent intensities were recorded for 10 s. Typically, the fluorescence signal photobleached to the background level in a single step during this 10 s period. For occasional double bleaches, presumably arising from two nearby labeled monomers, only the single fluorophore region before the final bleach was used for analysis.

The orientation of the probe was estimated by fitting analytical equations that predict the polarized fluorescence intensities during each complete illumination cycle as described in (19) and expanded here to include the $L$- and $R$- incident polarizations. Briefly, the raw intensity traces are corrected for instrument factors using a calibration procedure ((19) with additional terms in Appendix B) and the background is subtracted. Sixteen intensities (one complete 80 ms cycle) are combined in a mathematical model of the probe that describes its 3D orientation and rotational motions (see

Fig. **3**$A$) by approximating it as an electromagnetic dipole that absorbs and emits photons preferentially polarized along its dipole axis; see section A of the supplementary material for details. Corrections for partial mixing of the components of polarized fluorescent emission due to refraction by the high NA objective are included in the model; see (19, 22) and section E of the supplementary material for details.

The maximum likelihood values for the probe's 3D orientation and rotational wobble ($\theta$, $\phi$, $\delta$, $\kappa$) are determined for each 80 ms cycle using a Levenberg-Marquardt C algorithm that matches the predicted and measured intensities. $\kappa$ is an amplitude factor that scales the total intensity of the probe to the number of photons collected during each cycle. $\delta$ describes wobble motions on a time scale faster than measurement time (~10 ms) but slower than the fluorescent lifetime of the probe (~ 4 ns). The parameter describing rotational wobble on time scales faster than the fluorescent lifetime, $\delta_f$, was fixed during the analysis at 22.5° (15). The angular values ($\theta$,$\phi$) in the microscope coordinate frame were rotated into the coordinate frame of the actin filament ($\beta$,$\alpha$) using the direction of probe motion from the preceding pair of video frames as the polar axis in the actin frame of reference (see

Fig. **3***B*, Appendix C, and (22) for more details).

## Filament Length

In order to estimate the length of the actin filament, the sample was briefly illuminated with a HeNe laser to excite the Alexa 647 labeled phalloidin. The fluorescent emission was passed through a band pass filter (Omega 670DF40) and captured in a single image by the CCD. The Alexa 647 image of the filaments was overlaid with two subsequently acquired images of the rhodamine fluorophores to aid in molecule selection. Switching lasers required approximately 1-2 s so that matching the rhodamine actin with its filament was unambiguous. Filament lengths were measurable for the first 20 min of data recording. The length of the filament was estimated in ImageJ (30) by summing short line segments manually selected along the filament contour. Filament length could only be determined at low (0.03 and 0.1 mg/ml) myosin concentration due to shearing of the filaments, which for high (1 and 3 mg/ml) concentrations occurred within the first 1-2 min of adding motility buffer to the flow cell.

## RESULTS

### Myosin II

Polarized fluorescence intensities from a filament twirling about its axis during translocation show prominent oscillations (e.g., Fig. 2*A*). When all of the separate polarized fluorescence intensities are summed together the resulting total is a constant intensity that bleaches to background in a single step (see Fig. 2 in the supplemental material). Strong variation of the polarized fluorescence intensities with constant total intensity is an indicator of probe rotation. The intensities of the different channels oscillate with different phases as the rotating probe temporarily comes into alignment with the polarization direction of the incident illumination. Prior to maximum likelihood analysis, a 5-point running average is applied to the data (solid lines in Fig. 2*B*) in order to remove some high-frequency noise. The traces and their oscillations due to changes in $\alpha$ as the filament translocates are reproduced by the predictions of the fitted model (dotted lines in Fig. 2*B*).

In a plot of $\theta$ vs. $\phi$, actin filaments translocating uniformly (i.e., with constant velocity and direction within the *x-y* plane) that are also twirling with a constant angular velocity show a circular pattern that is centered at (90°, $\phi_{actin}$), see (Fig. 4 *A* and *B*). $\phi_{actin}$ is the direction in the *x-y* plane of the filament trajectory with respect to the +*x* axis. $\theta$ oscillates about 90° because a probe that is fixed in a twirling filament will spend half of each rotation pointing above the *x-y* plane and the other half pointing below it.

For a filament that is uniformly twirling about its axis, $\beta$ is approximately constant between 0° and 90° (Fig. 4*C*), and $\alpha$ increases (right-handed pitch) or decreases (left-handed pitch) linearly in the range -180° < $\alpha$ < 180° (Fig. 4*D*). The wobble, $\delta$, of the probe attached to actin is noisy (Fig. 4*E*), similar to results from (15), but it neither exhibits systematic variation with $\alpha$ and $\beta$, nor prevents determination of the probe orientation, as would be the case if the probe could freely rotate about its attachment point, i.e., $\delta$ = 90°. A continuous line for $\alpha$ can then be obtained by shifting the values after each full rotation (Fig. 4*F*) by 360°.

The angular frequency $\omega$ of filament twirling is calculated from the slope of the line fit to $\alpha$ *vs.* time. For the filament in Fig. 2, $\omega$ was 0.94 rotations/s and its linear velocity was 0.24 μm/s. The direction of rotation for $\alpha$ combined with the direction of motion of the filament, allows for determination of the handedness of actin twirling. To be sure of this handedness, we carefully checked the 3D orientation of the fluorescence signals with polarizers that extinguish the various polarized incident beams and the sign of the *x-y* coordinates of the camera images by physically moving the stage while observing the video output. The value of the twirling motion pitch is given by the ratio of the linear velocity to angular frequency ($v/\omega$), which for the filament of Fig. 2 is 0.25 μm. Simulated polarization intensities of a uniformly twirling filament with parameters similar to those in Fig. 4 ($\beta = \delta = 45°$ and $\Delta\alpha = 3$ rotations) are shown for comparison to the data in Fig. 1 of the supplemental material. The traces in Fig. 2 agree quite well with the prediction of the simulations despite the simplifications of constant $\beta$ and $\delta$.

Twirling assays were performed over a range of myosin and MgATP concentrations; however, most data were collected at 10 μM MgATP and 0.1 mg/ml myosin, where filament length was also determined. Recordings were selected for orientation analysis when the total intensity (i.e., the sum over both APD signals for each complete cycle plotted as a function of time, e.g., Fig. 2 of the supplemental material) prior to irreversible fluorophore photobleaching was $1.5 - 2 \times$ the total background intensity (711 out of 3,528 recordings). For further analysis, the filaments were required to meet several other criteria: (i) approximately constant rotational velocity (i.e., linear fit of $\alpha(t)$ vs. time with |correlation coefficient| > 0.9), (ii) a minimum total rotation about the filament axis of 180°, and (iii) a recorded signal duration before bleaching of at least 1.6 s. 144 filaments satisfied these criteria. After checking the 3D orientation results and fitted lines for these filaments, an additional 47 filaments were rejected due to apparent large, abrupt changes in $\theta$, $\phi$, or $\beta$ that were inconsistent with a smoothly twirling filament (see Fig. 3 and 4 of the supplemental material for examples and more detail on the selection process). Of the remaining 97 filaments, 94 and 3 of the filaments twirled with left- and right- handed pitches, respectively. There was no apparent trend or additional defining characteristic of the rejected filaments. Approximately 450-550 of the filaments had sufficient duration and signal quality but did not twirl, thus resulting in a twirling fraction of ~20%. Nearly all (>95%) of the filaments that twirled did so with a left-handed pitch, which is opposite to the intrinsic right-handed pitch of actin.

Excluding as outliers the three filaments that seemingly twirled with opposite pitch, the average twirling pitch of -0.47 ± 0.2 μm (mean ± SD, n = 94) is quite long compared to that of the long-pitch actin strands (74-76 nm). The angular velocity of the filaments were correlated with their linear velocity (Fig. 5). If the path of a twirling filament was dictated by structural considerations, then the linear and angular velocities would be coupled together in direct proportion. The lines fitted to the data at each myosin concentration, however, do not go through zero (angular and linear velocity are not directly proportional), suggesting that kinetic aspects determine the pitch as well as structural ones.

As myosin concentration loaded into the flow cell increased from 0.03 to 3.2 mg/ml, the velocity of the filaments increased for a fixed MgATP concentration (see Fig. 5 of the supplemental material). Obtaining polarization data from filaments moving faster than ~0.25 μm/sec (corresponding to >15 μM MgATP and > ~1 mg/ml myosin concentration), however, was difficult using the present polTIRF setup. In order to maximize the time traversing the relatively small APDs, lower MgATP concentrations were used at the higher myosin concentrations in most of the experiments. As a result, when the data from Fig. 5 were averaged and considered versus myosin concentration, the linear velocity (Fig. 6*A*) and pitch were fairly constant (Fig. 6*C*). A slight decrease of the magnitude of the (negative) angular frequency over this range (Fig. 6*B*) is not statistically significantly.

At 0.03 and 0.1 myosin concentrations, filament velocity is independent of filament length, see Fig 6 of the supplemental material, as reported previously (31). The twirling frequency and pitch are also insensitive to filament lengths longer than ~ 1 μm (Fig. 7).

## Myosin V

Most of the myosin V data was collected at 500 and 1000 μM MgATP, which corresponded to average velocities of 0.14 and 0.22 μm/s. 201 recordings were selected for analysis using criteria similar to the myosin II data. 158 filaments met the criteria but did not twirl whereas 43, or nearly 20%, showed clear twirling motions. All but one of the filaments exhibited a left-handed pitch with a magnitude of 1.7 ± 0.1 μm. Filament twirling by myosins II and V are summarized in Table 1 and also compared with myosin VI data as reported previously (32). Fluorescence polarization intensities and probe angles for a typical filament twirling by myosin V is shown in Fig. 7 of the supplemental material. The myosin V data were obtained while the setup presented in this paper was under development. Consequently, only two ±45° polarizations (L1 and R1) were available in beam 1, resulting in a total of 8 recorded fluorescence intensities. The angular resolution was limited to one quarter of a hemisphere by selecting only those filaments that translocated within ±15° of the $x$ axis, thereby restricting the probe angle in the laboratory coordinate frame to $-90° < \phi < 90°$ and allowing the handedness of uniformly twirling filaments to be determined.

## DISCUSSION

## Angular scope of polTIRF measurements

Several previous reports (12, 14-17) have indicated that under various conditions, myosin isoforms produce a torque either in conjunction with axial force production, or while stepping along actin in a helical path. Our previous work using polarized total internal fluorescence (polTIRF) microscopy (15) indicated filament twirling; however, symmetries in the measurement of the probe orientation resulted in an ambiguous sign for the azimuthal angle and thus did not report the handedness. These symmetries were broken here by incorporating incident polarizations ($L$ and $R$) that are intermediate between $p$ and $s$ into the polTIRF method (20). The only remaining ambiguity of probe orientation is unavoidable due to its inherent dipole symmetry: ($\theta, \phi$) is equivalent to ($\pi - \theta, \phi + \pi$).

Recordings of actin filaments, labeled with rhodamine at Cys[374], and gliding under the propulsion of myosin isoforms II or V (this work) or myosin VI (32) corresponded to oscillations of the probe orientation ($\theta, \phi$), which in the actin frame of reference were consistent with a fairly constant axial angle $\beta$ and linearly increasing or decreasing azimuthal $\alpha$. This angular response of the probe during motility shows that the filament often twirls about its axis during translocation, and also confirms that the method, with the additional input polarizations and the attendant analysis, resolves the probe orientation within a hemisphere as expected. Twirling generated by myosin II and V has a strong left-handed bias while twirling by myosin VI was strongly right-handed, arguing against a dominating experimental artifact, which would likely result in twirling with either handedness independent of the myosin isoform.

## Non-twirling filaments

Similar to earlier reports (16), only ~20% of the filaments translocated by myosin V showed clear twirling rotations. In our twirling assay with myosin II most of the filaments translocated, but only 20% clearly twirled, thus indicating that azimuthal rotation is not necessary for gliding. In this work, probe orientation and location were not measured simultaneously; consequently the initial direction of gliding was assumed to be maintained during the polarization analysis. If a filament were to undergo large, unobserved changes in this motion, such as stopping or turning, it would likely be rejected during screening of the filaments. Such motions, however, are unlikely to account for all of the non-twirling filaments since U-turns and stopping events were observed for only 10-20% of filaments during a 30 s observation, which is much less than the 80% of filaments that did not twirl. Non-twirling filaments were difficult to characterize, but they rarely exhibited steady $\alpha$ and $\beta$ orientations, which would be expected if the filament slides perfectly straight or stops moving completely, as is seen when ATP is omitted (15).

The suspended-filament bead assay reported earlier for myosin V (16) and VI (17) showed that often the motors stepped straight along the filament, rather than in a helical path. Thus, both twirling and non-twirling motions are observed in experiments with single motors and in gliding assays where multiple motors are involved.

## Handedness of twirling

Measuring the probe orientation unambiguously for half of a rotation or more allows the direction of angular rotation to be determined. Combining this angular direction with the direction of filament sliding indicates the handedness of actin filament twirling; a feature not obtained in previous experiments using single molecule polarized fluorescence (14, 15). We determined that when filaments twirl by fast skeletal whole myosin II and native myosin V from chick brain, they consistently have a left-handed pitch, opposite to the intrinsic right-handed long pitch of actin. Filaments twirled with a relatively long pitch of 0.47 and 1.7 µm for myosin II and V, respectively, see Table 1. We previously reported that myosin VI twirls filaments with the opposite, right-handed pitch (32). Right-handed twirling by myosin VI is not simply due to its 'backward' pointed-end directed motility, as can be realized by considering a nut twirling on a machine screw, where the handedness is the same for tightening and loosening. Preliminary twirling experiments using myosin V with a truncated lever arm, having only 4 calmodulin-binding IQ motifs, twirls a higher fraction of filaments to the right than full-length myosin V with 6 IQ-motifs per head (J. H. Lewis, personal communication). The truncated lever arm results suggest that the handedness is related to the step distance of these molecules.

Single and multiple molecules of myosin V move a proportion (20%) of bead duplexes in a helical path around a suspended actin filament (16) with the same left-handedness and a similar pitch (2.2-2.5 µm) as twirling filaments in our gliding assay, pitch ~1.7 µm. Similar experiments with myosin VI (17) gave a right handed pitch, in agreement with the polTIRF twirling assay (32), although the pitch of the bead-duplex path was longer (2.2-5.6 µm) than the twirling pitch (1.3 ±0.1 µm) in the gliding assay. Regardless of whether multiple motors, fixed on a microscope slide, were translocating actin, or a single motor was stepping along a suspended filament, the rotational motion is similar. This agreement in handedness between the two different assays suggests that a similar mechanism underlies torque generation in these two geometries.

Formation of actin filament superhelices by HMM translocating filaments at 1 mM MgATP (12) indicate a right-handed torque, opposite to that of the twirling actin filaments obtained here. This discrepancy might arise from the difference in MgATP concentration, which was higher in the superhelix experiments and limited in the twirling experiments to produce velocities compatible with time resolution and size of the photodetector in the present setup.

Another difference is that the supercoiling filaments (12) were specifically attached to the slide at their pointed ends and the remaining filament forced to buckle by a track of HMM, created by decorating the filament prior to surface attachment. Twirling filaments here were moved by myosin molecules randomly distributed on the surface.

## Possible Mechanisms of Twirling

The mechanisms responsible for twirling filaments are not known, but the experiments presented here suggest several possible effects: a torque component to the myosin power stroke (12), (16), the step size of myosin relative to the actin filament helix (14, 16), or the drag force of rigor heads on the filament. A torque between myosin and actin could arise if the vector of the myosin working stroke is not parallel to the axis of the actin filament, thus resulting in a small angular component to the velocity in addition to the linear component. An estimate of the vectorial direction of the working stroke for myosin II can be obtained from crystal structures in the pre-power stroke position (33) and near the end of the power stroke (34) when both structures are docked into cryo-electron microscopy density maps of actin decorated with myosin subfragment 1 (5). The flexible joint between the light chain domain and myosin rod is located near $Lys^{843}$ of chicken skeletal myosin. Between the pre-power stroke and the near rigor state of the docked structures, $Lys^{843}$ moves predominantly along the filament axis, but with a slight (~20º) right-handed tilt relative to the actin axis. A tilt in the other direction would be required to induce the typical twirling pitch we detected in myosin II and V. Thus the left-handed twirling we observed is not explained by the path of the power stroke, unless the docked structures are misoriented by at least 20°. Also, if the torque is generated directly by the working stroke, then a strong coupling between linear and angular velocities would be expected in the twirling assay. Our measurements show, however, only weak coupling between these two quantities.

Alternatively, the handedness seen in processive bead assays (16, 17) could arise from the interaction between the stride length of myosin and the helical disposition of the actin monomers. Actin filaments are torsionally flexible, but often approximated as a 13/6 helix when interpreting the path of myosins on actin (16, 17). The 13/6 index defines an actin filament with monomers disposed around the left-handed short-pitch 'genetic' helix rotating 6 full turns in 13 monomers. Starting from the zeroth actin monomer as the origin, the 13th monomer along this short pitch helix is located at the same azimuthal orientation around the filament as the original one. Each monomer is $360° \cdot (-6 / 13) \approx -166°$ azimuthally around the short pitch helix (negative sign means to the left). The $2^{nd}$ monomer is positioned at: $-166° \cdot 2 = -332° = 360-332 = 28°$, i.e., on the right-handed long-pitch strand. On the 13/6 helix, a motor whose stride is 13 monomers (~36 nm) would walk straight. Left-handed rotation would arise if the motor made more 11 monomer steps than 15, and *vice versa* for a right-handed helical path.

Cryo-EM data showing the spacing of myosin V rigor heads bound to actin (35) indicated a distribution of stride lengths spanning mostly 13 monomers, and significant amounts of 11 and 15 monomer spans. There were more 15s than 11s, which would imply a bias toward a right-handed helical path. Our twirling results and the experiments with suspended actin filaments(16, 17), however, show left-handed bias.

Reference (36) pointed out that this apparent discrepancy would be resolved if the filament structure is not a 13/6 helix. For instance a 28/13 helix, which has been also observed (37) has an azimuthal rotation of the left-handed genetic helix of $360° \times (-13/ 28) \approx -167°$. The $13^{th}$ and $15^{th}$ actin monomers are positioned at 13° to the left and right of the axial (straight) direction, respectively. Then a motor, such as myosin V, which takes more 13-monomer steps than 15s (as shown in the EM distribution of spans (35)) would walk slightly left-handed, as we observe.

Myosin VI adopts a straight or slightly right-handed helical path (17, 32). This was interpreted as indicating an even longer stride than myosin V(16), but it could also indicate the presence of much shorter steps, for instance the proper ratio of 6 and 7 monomer steps on a 13/6 helix (azimuthally oriented at 83° to the left and right, respectively) could also result in a slight right-handed helical pitch. Myosin VI has a very wide distribution of step sizes (32, 38), making the azimuthal path of the individual molecules highly variable (32).

It is less clear how to extend the effect of stride-length to the smaller, non-processive steps of myosin II. If they followed the long-pitch actin helix, then myosin II would twirl actin with a short right-handed pitch (74 nm) and not the more gradual left-handed motion that is observed. The step-size idea also predicts a strong correlation between linear velocity and azimuthal twirling frequency, variables that are only weakly correlated in the data here (Fig. 5).

An interesting suggestion (A. Vilfan, personal communication) is that left-handed actin twirling by non-processive myosin in a gliding assay arises from myosin heads preferentially binding to actin monomers that approach at the correct azimuth. This preference arises because only a fraction of the monomers, which are helically distributed around the filament, lie within a range of azimuthal orientations suitable for binding of myosin ("target zone" (39)). These monomers are presented to the head at regular intervals as the filament translocates above it. In the 13/6 actin helix, if the myosin head binds to the 13 monomer, then the filament would glide straight (black arrow in Fig. 8A). If attachment is very fast, however, then monomers within the range of acceptable azimuths (e.g., the hatched area in Fig. 8A) and leading the center of the target zone (e.g., number 11) will become the preferred binding site. In a gliding assay the torque arises as the left-of-center monomer (monomer 11) is pulled in toward the filament axis during the stroke (red arrow) analogous to the mechanism proposed for the processive bead assays (16, 17). A prediction of this model is that an experiment at higher ionic strength, which would weaken myosin binding to actin, might show straighter or even right-handed twirling. Higher velocity filaments, due to higher MgATP concentrations, might also affect the twirling pitch by altering which monomers within the target zone are the most likely binding sites.

Finally, we consider the drag force of the rigor heads, which are bound tightly to actin, and are probably the dominant retarding force on a gliding actin filament at the low MgATP concentrations used here. If forward motion of the filament is due to a nearly straight or slightly right-handed power stroke (12), then left-handed twirling may result from a drag force due to rigor heads pulling backwards but also slightly laterally. Hopkins et al. (40) measured tilting and twisting of myosin regulatory light chains in muscle fibers. They reported a twisting motion of the myosin (i.e., the $\gamma$ angle) when a rigor muscle is allowed to shorten, but no twisting in an active muscle. The analogous drag force of rigor heads in a gliding assay would result in a left-handed torque to the gliding filament. In Fig. 8B, the resultant velocity (red arrow) on an actin filament is shown as the vector sum of velocities due to the power stroke (black arrow) from active heads and drag from strongly attached rigor heads (blue arrow). The rigor-drag hypothesis may explain the insensitivity of twirling to filament length and myosin concentration because twirling would instead be attributed to the ratio of active and rigor heads along the filament and not their absolute number. Unlike models that explain twirling via a direct structural coupling between linear and angular velocity, the rigor-drag hypothesis would predict a larger pitch at higher MgATP concentrations because linear velocity would increase as angular velocity decreased (since there are fewer rigor heads bound to actin). This prediction has not been tested critically in this work or previous experiments (14) due to the limited range of MgATP concentrations tested. Since the fraction of heads bound to a filament in the rigor state decreases at high MgATP concentrations, the tendency to twirl with a left-handed pitch would be reduced, and thus possibly reconcile the difference between the left-handed twirling at low MgATP concentrations measured in this work with the experiments by Nishizaka et al. (12) that showed a right-handed torque at high (2 mM) MgATP concentration.

## CONCLUSIONS

PolTIRF has been extended to enable measurement of probe angles within a hemisphere of solid angle. It has been used to determine the handedness of twirling actin filaments gliding over a surface containing myosin II or V. Approximately 20% of analyzable filaments, translocated by myosin II or V, twirled with left-handed pitch. The magnitude of the pitch depends on myosin isoform, but for myosin II is insensitive to filament velocity, filament length or myosin concentration in the range investigated. A torque component to the working stroke, the myosin stride length, the helical distribution actin monomers, and the drag of rigor myosin heads may contribute to the azimuthal force component in a gliding assay, but no single one of these influences adequately describes the rotational motions observed.

## SUPPLEMENTAL MATERIAL

An online supplement to this article can be BJ Online at http://www.biophysj.org.

We would like to thank A. Gilmour, R. Kuduravalli, T. D. Logan, T. Lin, T. Liu, M.L. Ohlmann, E.M. Ostap, H. Pham, Y. Sun, K.B. Towles, and A. Vilfan for technical assistance and helpful suggestions. PCN and JFB were supported in part by NSF grants DGE-0221664, DMR04-25780, and DMR-0404674. YEG was supported in part by DMR04-25780 and R01-AR026846. HWS was supported by P01AR051174-05.

## REFERENCES

1. Cooke, R. 1997. Actomyosin interaction in striated muscle. Physiol Rev 77:671-697.
2. Rayment, I., H. M. Holden, M. Whittaker, C. B. Yohn, M. Lorenz, K. C. Holmes, and R. A. Milligan. 1993. Structure of the actin-myosin complex and its implications for muscle contraction. Science 261:58-65.
3. Geeves, M. A., and K. C. Holmes. 2005. The molecular mechanism of muscle contraction. Adv Protein Chem 71:161-193.
4. Warshaw, D. M. 2004. Lever arms and necks: a common mechanistic theme across the myosin superfamily. J Muscle Res Cell Motil 25:467-474.
5. Dobbie, I., M. Linari, G. Piazzesi, M. Reconditi, N. Koubassova, M. A. Ferenczi, V. Lombardi, and M. Irving. 1998. Elastic bending and active tilting of myosin heads during muscle contraction. Nature 396:383-387.
6. Gillis, J. M., and E. J. O'Brien. 1975. The effect of calcium ions on the structure of reconstituted muscle thin filaments. J Mol Biol 99:445-459.
7. Maeda, Y., I. Matsubara, and N. Yagi. 1979. Structural changes in thin filaments of crab striated muscle. J Mol Biol 127:191-201.

8.      Toprak, E., J. Enderlein, S. Syed, S. A. McKinney, R. G. Petschek, T. Ha, Y. E. Goldman, and P. R. Selvin. 2006. Defocused orientation and position imaging (DOPI) of myosin V. Proc Natl Acad Sci U S A 103:6495-6499.

9.      Syed, S., G. E. Snyder, C. Franzini-Armstrong, P. R. Selvin, and Y. E. Goldman. 2006. Adaptability of myosin V studied by simultaneous detection of position and orientation. Embo J 25:1795-1803.

10.     Ross, J. L., M. Y. Ali, and D. M. Warshaw. 2008. Cargo transport: molecular motors navigate a coplex cytoskeleton. Currrent Opinion in Cell Biology 20.

11.     Matsubara, I., Y. E. Goldman, and R. M. Simmons. 1984. Changes in the lateral filament spacing of skinned muscle fibres when cross-bridges attach. Journal of Molecular Biology 173:15-33.

12.     Nishizaka, T., T. Yagi, Y. Tanaka, and S. Ishiwata. 1993. Right-handed rotation of an actin filament in an in vitro motile system. Nature 361:269-271.

13.     Suzuki, N., H. Miyata, S. Ishiwata, and K. Kinosita, Jr. 1996. Preparation of bead-tailed actin filaments: estimation of the torque produced by the sliding force in an in vitro motility assay. Biophys J 70:401-408.

14.     Sase, I., H. Miyata, S. Ishiwata, and K. Kinosita, Jr. 1997. Axial rotation of sliding actin filaments revealed by single-fluorophore imaging. Proc Natl Acad Sci U S A 94:5646-5650.

15.     Rosenberg, S. A., M. E. Quinlan, J. N. Forkey, and Y. E. Goldman. 2005. Rotational motions of macro-molecules by single-molecule fluorescence microscopy. Acc Chem Res 38:583-593.

16.     Ali, M. Y., S. Uemura, K. Adachi, H. Itoh, K. Kinosita, Jr., and S. Ishiwata. 2002. Myosin V is a left-handed spiral motor on the right-handed actin helix. Nat Struct Biol 9:464-467.

17.     Ali, M. Y., K. Homma, A. H. Iwane, K. Adachi, H. Itoh, K. Kinosita, Jr., T. Yanagida, and M. Ikebe. 2004. Unconstrained steps of myosin VI appear longest among known molecular motors. Biophys J 86:3804-3810.

18.     Quinlan, M. E., J. N. Forkey, and Y. E. Goldman. 2005. Orientation of the myosin light chain region by single molecule total internal reflection fluorescence polarization microscopy. Biophys J 89:1132-1142.

19.     Forkey, J. N., M. E. Quinlan, and Y. E. Goldman. 2005. Measurement of single macromolecule orientation by total internal reflection fluorescence polarization microscopy. Biophysical Journal 89:1261-1271.

20.     Prummer, M., B. Sick, B. Hecht, and U. P. Wild. 2003. Three-dimensional optical polarization tomography of single molecules. Journal of Chemical Physics 118:9824-9829.

21.     Forkey, J. N., M. E. Quinlan, M. A. Shaw, J. E. Corrie, and Y. E. Goldman. 2003. Three-dimensional structural dynamics of myosin V by single-molecule fluorescence polarization. Nature 422:399-404.

22.     Beausang, J. F., Y. Sun, M. E. Quinlan, J. N. Forkey, and Y. E. Goldman. 2008. Orientation and Rotational Motions of Single Molecules by Polarized Total Internal Reflection Fluorescence Microscopy. In Single Molecule Techniques. P. R. Selvin, and T. Ha, editors. Cold Spring Harbor Laboratory Press, Cold Spring Harbor. 121-148.

23. Beausang, J. F., J. H. Lewis, W. H. Schroeder 3rd, H. L. Sweeney, and Y. E. Goldman. 2008. 2252-Pos Twirling Of Actin By Myosin Isoforms. Biophysical Journal 94:2252-Pos.

24. Margossian, S. S., and S. Lowey. 1982. Preparation of Myosin and Its Subfragments from Rabbit Skeletal-Muscle. Methods in Enzymology 85:55-71.

25. Cheney, R. E. 1998. Purification and assay of myosin V. Methods Enzymology 298:3-18.

26. Spudich, J. A., and S. Watt. 1971. Regulation of Rabbit Skeletal Muscle Contraction .1. Biochemical Studies of Interaction of Tropomyosin-Troponin Complex with Actin and Proteolytic Fragments of Myosin. Journal of Biological Chemistry 246:4866-&.

27. Corrie, J. E., and J. S. Craik. 1994. Synthesis and characterization of pure isomers of iodoacetamido-tetramethylrhodamine. J Chem Soc Perkin Trans 1:2967-2974.

28. Putkey, J. A., K. F. Ts'ui, T. Tanaka, L. Lagace, J. P. Stein, E. C. Lai, and A. R. Means. 1983. Chicken calmodulin genes. A species comparison of cDNA sequences and isolation of a genomic clone. Journal of Biological Chemistry 258:11864-11870.

29. Press, W. H., S. A. Teukolsky, W. T. Vetterling, and B. P. Flannery. 1992. Numerical Recipes in C. Cambridge University Press, Cambridge.

30. Rasband, W. S. 1997-2007. ImageJ. U.S. National Institutes of Health, Bethesda, MD. http://rsb.info.nih.gov/ij/.

31. Uyeda, T. Q., S. J. Kron, and J. A. Spudich. 1990. Myosin step size. Estimation from slow sliding movement of actin over low densities of heavy meromyosin. J Mol Biol 214:699-710.

32. Sun, Y., H. W. Schroeder, 3rd, J. F. Beausang, K. Homma, M. Ikebe, and Y. E. Goldman. 2007. Myosin VI walks "wiggly" on actin with large and variable tilting. Mol Cell 28:954-964.

33. Dominguez, R., Y. Freyzon, K. M. Trybus, and C. Cohen. 1998. Crystal structure of a vertebrate smooth muscle myosin motor domain and its complex with the essential light chain: Visualization of the pre-power stroke state. Cell 94:559-571.

34. Rayment, I., W. R. Rypniewski, K. Schmidt-Base, R. Smith, D. R. Tomchick, M. M. Benning, D. A. Winkelmann, G. Wesenberg, and H. M. Holden. 1993. Three-dimensional structure of myosin subfragment-1: a molecular motor. Science 261:50-58.

35. Walker, M. L., S. A. Burgess, J. R. Sellers, F. Wang, J. A. Hammer, J. Trinick, and P. J. Knight. 2000. Two-headed binding of a processive myosin to F-actin. Nature 405:804.

36. Vilfan, A. 2005. Influence of fluctuations in actin structure on myosin V step size. Journal of Chemical Information and Modeling 45:1672-1675.

37. Yagi, N., and I. Matsubara. 1989. Structural changes in the thin filament during activation studied by X-ray Diffraction of Highly Stretched Skeletal Muscle. J Mol Biol 208:359-363.

38. Yildiz, A., H. Park, D. Safer, Z. Yang, L.-Q. Chen, P. R. Selvin, and H. L. Sweeney. 2004. Myosin VI steps via a hand-over-hand mechanism with its lever arm undergoing fluctuations when attached to actin. Journal of Biological Chemistry 279:37223-37226.

39. Taylor, K. A., M. C. Reedy, L. Cordova, and M. K. Reedy. 1984. Three-dimensional reconstruction of rigor insect flight muscle from tilted thin sections. Nature 310:285-291.

40. Hopkins, S. C., C. Sabido-David, U. A. van der Heide, R. E. Ferguson, B. D. Brandmeier, R. E. Dale, J. Kendrick-Jones, J. E. Corrie, D. R. Trentham, M. Irving, and Y. E. Goldman. 2002. Orientation changes of the myosin light chain domain during filament sliding in active and rigor muscle. J Mol Biol 318:1275-1291.

**TABLE**

| Myosin Isoform | II | V | VI |
|---|---|---|---|
| Myosin, mg/ml | 0.1-3.3 | 0.26 | 0.26 |
| Ave. Velocity, μm/s | 0.091- 0.61 | 0.07 - 0.27 | 0.09 - 0.21 |
| ATP, μM | 5 - 20 | 10 - 2000 | 500 - 1000 |
| Handedness | Left | Left | Right |
| Number | 94 | 42 | 23 |
| Frequency, rev/s | -0.50 ± 0.19 | -0.12 ± 0.12 | 0.14 ± 0.05 |
| Total Rotation αΔ, rev | -1.4 ± 0.7 | -0.6 ± 0.4 | 0.7 ± 0.4 |
| Pitch, μm | -0.47 ± 0.19 | -1.7 ± 0.6 | 1.3 ± 0.5 |
| Probe Angle β, ° | 49 ± 19 | 55 ± 12 | 54 ± 9 |
| Slow Wobble δ, ° | 48 ± 17 | 30 ± 13 | 30 ± 11 |

Summary of actin filament twirling by myosins II, V and VI (mean ± SD). The range of angular rotations (αΔ) for myosins II, V and VI are -0.5 to -4.4, -0.09 to -1.8, and +0.2 to +2.0 rotations, respectively. 'Number' indicates those molecules twirling with the dominant handedness, which are the basis for the other average values in the table (see text for details). A minority of filaments (3, 1, and 0 for myosins II, V, and VI, respectively) twirled with the opposite handedness, and are not included in the table averages.

**FIGURES**

Fig. 1

At the sample plane, the two incident laser beams alternately propagate nearly along the x (beam 1) and y (beam 2) axes, illuminating the sample with one of four linear polarizations: s, p, L, or R for each path (see inset). For clarity, a portion of beam 2 is omitted. It reflects from the sample interface similar to beam 1, but propagates in the y-z plane perpendicular to the page. Beam 1 is focused by a lens (L1) at a glancing angle through a coupling prism (CP) onto the sample (SAM) and terminated in a beam dump (BD). Reflecting mirrors are shown as thick black lines. The 532 laser intensity is controlled by a rotatable half wave plate (HWP) and vertical polarizer (P0). The polarization of the beam is switched between vertical and horizontal polarizations by Pockels cell (PC0), passed through a shutter (S) and directed along paths 1 and 2 by a polarizing beam splitter (PBS0) and clean-up polarizers (P1 and P2). The polarization in each path is controlled by a Berek compensator (BC1 or BC2) and a second Pockels cell (PC1 or PC2), which change the polarization every 10 ms to $p$, $s$, $L$ and $R$. The emitted fluorescence intensity is collected by a microscope objective lens, passed through a barrier filter (BF) and directed onto either a CCD camera or reflected by a removable mirror (RM) onto two avalanche photodiodes (APDx and APDy) that measure the fluorescent intensity polarized along the $x$ and $y$ axes by PBS1. A small fraction of the beam prior to the shutter (S) is directed into a feedback circuit (see (19)) that controls the polarization from PC0 in order to maintain high extinction between beams 1 and 2.

Fig. 2
Typical polarized fluorescence intensity data for a single molecule of rhodamine attached to a twirling actin filament with 0.1 mg/ml myosin loading concentration, 10 μM MgATP, and velocity of 0.24 μm/s along the $x$-axis. The filament translocated at an angle of -95° (i.e, $\phi_{actin}$) relative to the positive $x$-axis. (A) Calibration factors for the 16 recorded intensities have been applied to the raw data and the background of each has been subtracted. (B) The data prior to the bleach at ~3.6 s (black triangle, see Fig. 2 of the supplemental material for the total fluorescence intensity) is passed through a 5-point (0.4 s) mean filter before maximum likelihood analysis for extracting angles. The resulting intensity calculated from the probe model is in good agreement (dotted) with the filtered data (solid). Peak intensities occur when the probe is aligned with the detector and the incident polarization. Consequently, evidence for twirling of the probe can be directly seen in the data as sets of oscillating intensity curves that are out of phase with one another. For example, in this case the data recorded for Beam 1 by APD y shows a pattern of sequential peak intensities $_{R1}I_y$, $_{s1}I_y$, $_{L1}I_y$ that repeats with each rotation of the probe (see also Fig. 1 of the supplemental material).

Fig. 3
Calculations are performed in the microscope coordinate frame (A), where the $+z$ axis is the optical axis, pointing toward the microscope objective, and the $+x$ and $+y$ axes are in the plane of the quartz/water interface and aligned with the plane of propagation of beams 1 and 2, respectively. The evanescent TIRF field decays in the $+z$ direction. $\theta$ is the polar angle of the probe with respect to the $+z$ axis and $+\phi$ is the azimuthal angle of the probe around $z$ defined positively from the $+x$ axis towards the $+y$ axis. The extent of fast ($\delta_f$, not shown) and slow wobble ($\delta$) motions are included in the analysis model to account for mixing between the polarizations due to motions of the probe on two time scales, both much faster than the twirling rate. Twirling is quantified in a frame of reference relative to the actin filament (B) where $\beta$ is the polar angle of the probe with respect to the forward moving end of the actin filament axis and $\alpha$ is the azimuthal angle around the filament axis.

Fig. 4
For each complete cycle of the polarization data in Fig. 2, the maximum likelihood angle in the microscope frame ($\theta, \phi$) and actin frame ($\beta, \alpha$) are plotted. Twirling is seen in the microscope frame either (A) as oscillations in time of $\theta$ and $\phi$ or (B) as a circle when plotted as $\theta$ vs. $\phi$ (open circle is t = 0 s). In the actin frame twirling is indicated by (C) a relatively constant $\beta$, here equal to ≈ 40°, and (D) saw-tooth shape for $\alpha$(t). (F) When $\alpha$(t) is shifted by 360° after each rotation, the saw-tooth becomes a straight line with slope $\omega$ equal to the twirling frequency, here ≈ -0.94 rev/s. The direction of translocation (along the $+x$-axis) is required for determining the handedness of the twirling motion; left- and right-handed are depicted here by a negative and positive slope for $\alpha(t)$, respectively. The slow wobble cone ($\delta$) of the probe is ~ 45°. Approximately half of the twirling filaments have traces similar to the quality shown here.

Fig. 5
Angular frequency versus average filament velocity with linear fits (solid) and 95% confidence interval (*dashed*) for the fitted lines for (A) 0.03, (B) 0.1, (C), 1.0, (D) 3.2 mg/ml myosin II loading concentration. Filled points correspond to left-handed twirlers (negative angular frequency). Open points, which are not included in the fit, correspond to right-handed twirlers (positive angular frequency).

Fig. 6
Average values and linear fits (*solid*) for (A) velocity, (B) angular frequency, and (C) pitch computed at each myosin concentration (from the data in Fig. 5) with error bars determined by the bootstrap method (29). Negative values of angular frequency indicate left-handed rotation in the actin reference frame. The magnitude of the twirling frequency decreased with increasing myosin concentration while the velocity of the filaments was relatively constant. Predictions (*dashed*) in each panel are calculated from the fits of the other two panels using the relation $\lambda = v/\omega$ and are in approximate agreement with their respective fits.

Fig. 7
Average (A) velocity, (B) angular frequency, and (C) pitch as a function of filament length at 0.03 (*gray*) and 0.1 mg/ml (*black*) loading concentration of myosin with linear fits (*solid*) and 95% confidence interval of the fit (*dashed*).

Fig. 8
Two models for actin twirling: (A) Target zone model. Monomer 11 of the translocating actin filament (*gray cylinder*, pointed-end up) enters the target zone (*shaded*) of the myosin head (*black dot*) before monomer 13 and therefore binds more rapidly to myosin, despite lying slightly off the filament axis. Twirling of the filament occurs during the power stroke as monomer 11 is pulled predominantly along the direction of filament motion, but also in toward the filament axis (*red arrow*). Myosin attachment to on-axis monomer 13, however, would translocate actin during the power stroke without twirling the filament (*black arrow*). For convenience, monomers are labeled relative to a supposed previous binding site one helical repeat away at monomer 0 along the 13/6 actin helix. (B) Rigor drag model: Hypothetical motions for myosin attached to actin are shown by arrows at a possible binding site along the actin helix. The motion due to the power stroke of the myosin heads attached to actin (*solid black*) is mostly along the $z$-direction, but may also apply a slightly right-handed torque (5, 12). The drag due to the rigor heads attached to actin (*blue*) retards the linear motion, but instead of acting along the same direction as the power stroke (*dotted black*), it is sufficiently skewed to cause the resultant motion (*red*) to be slightly left-handed.

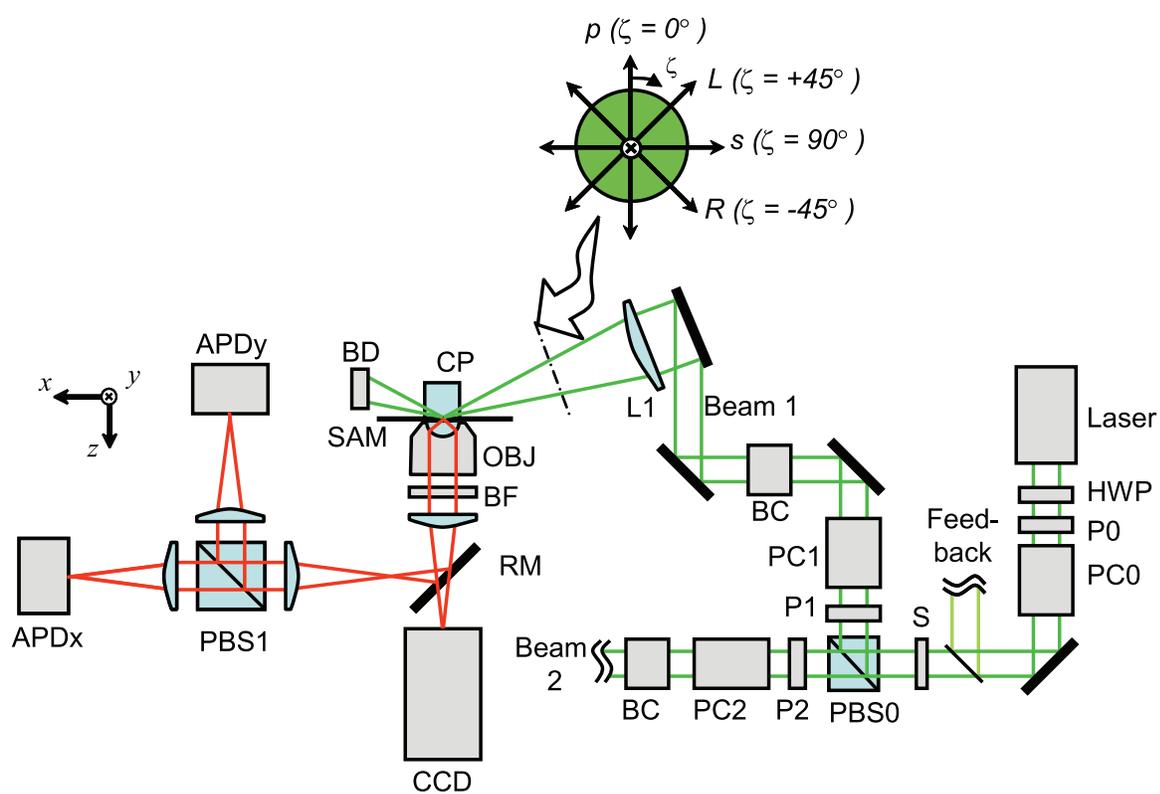

Figure 1:

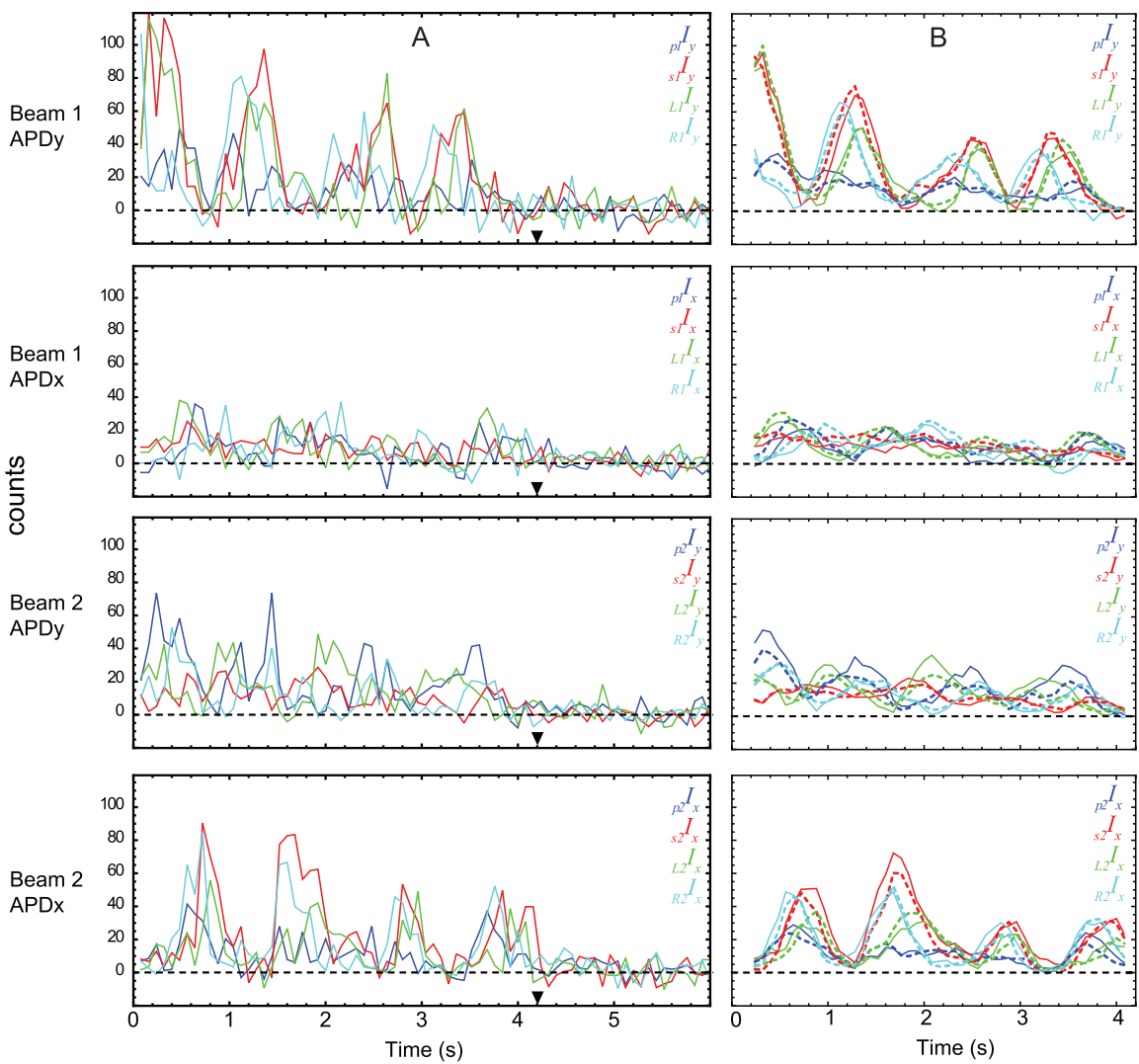

Figure 2:

A

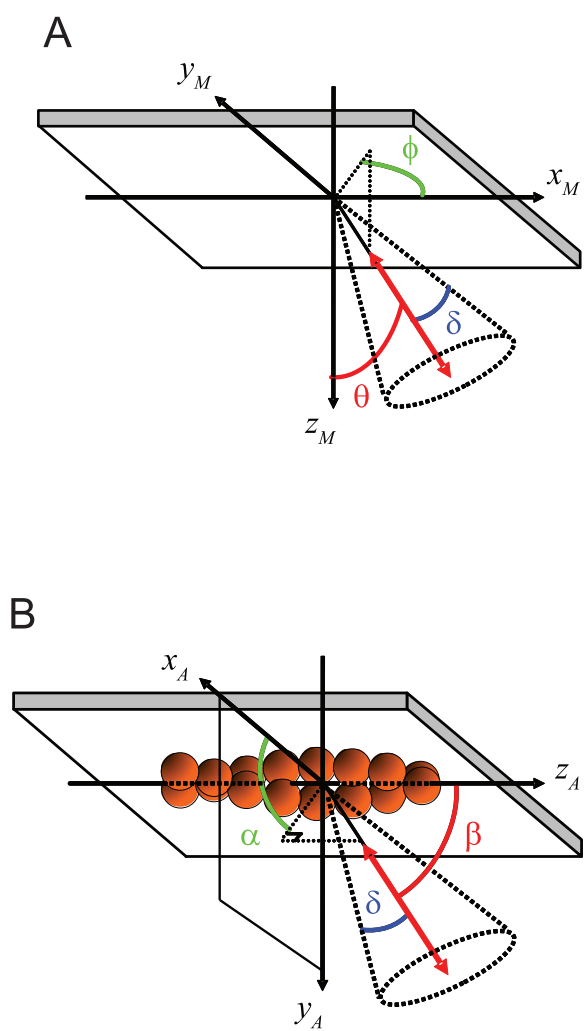

B

Figure 3:

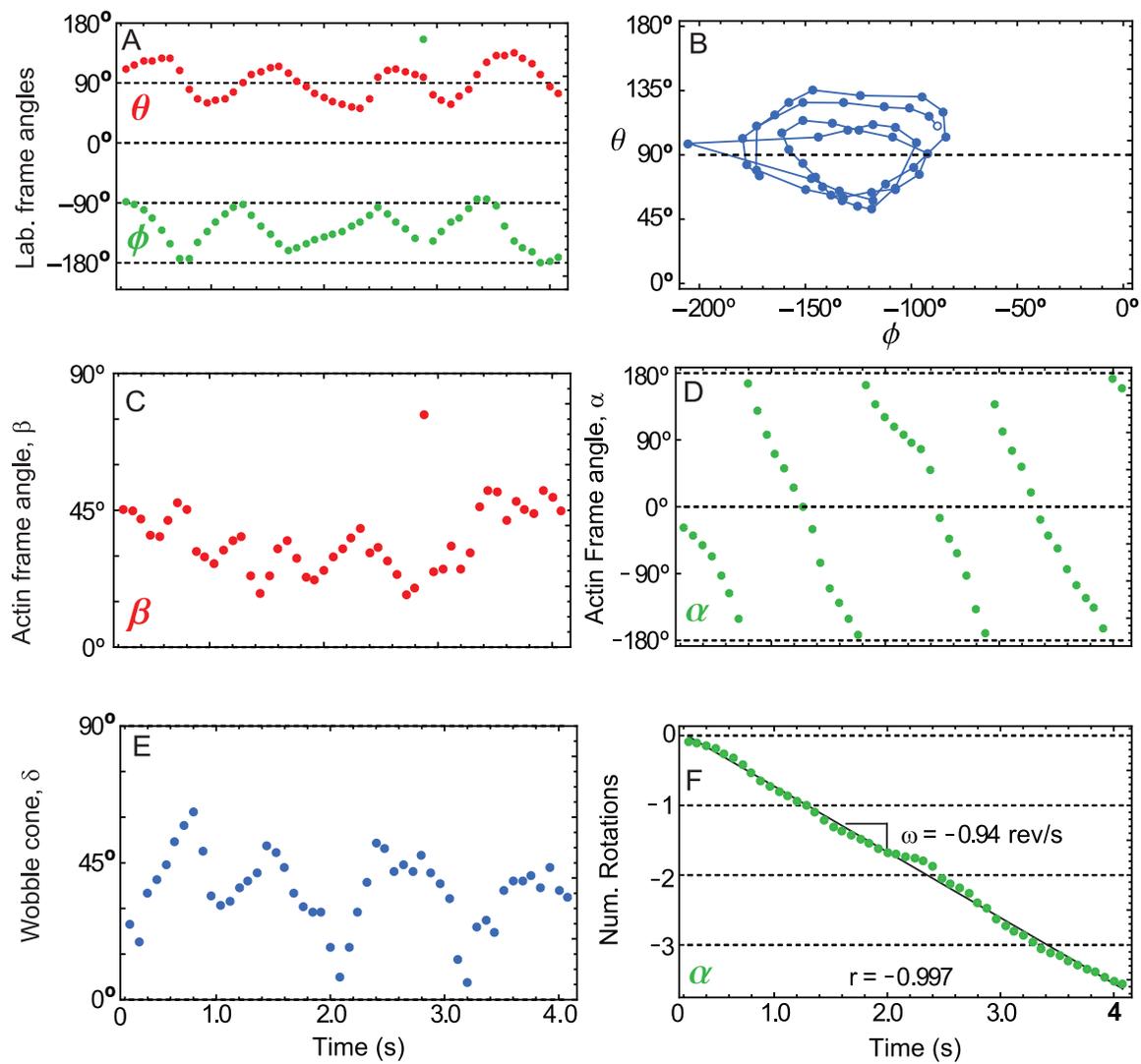

Figure 4:

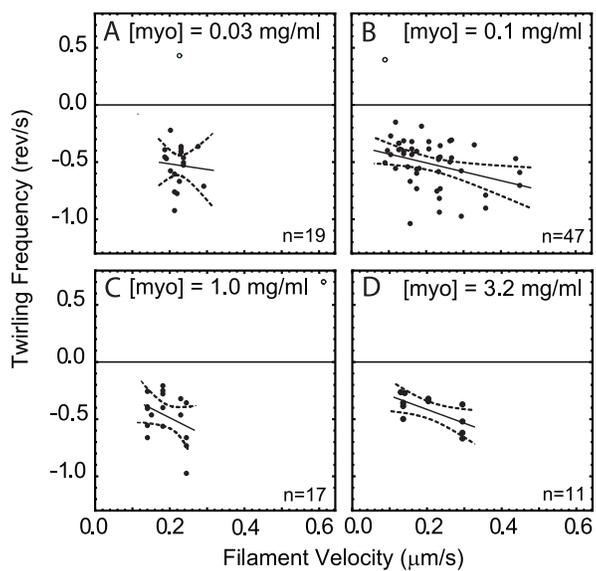

Figure 5:

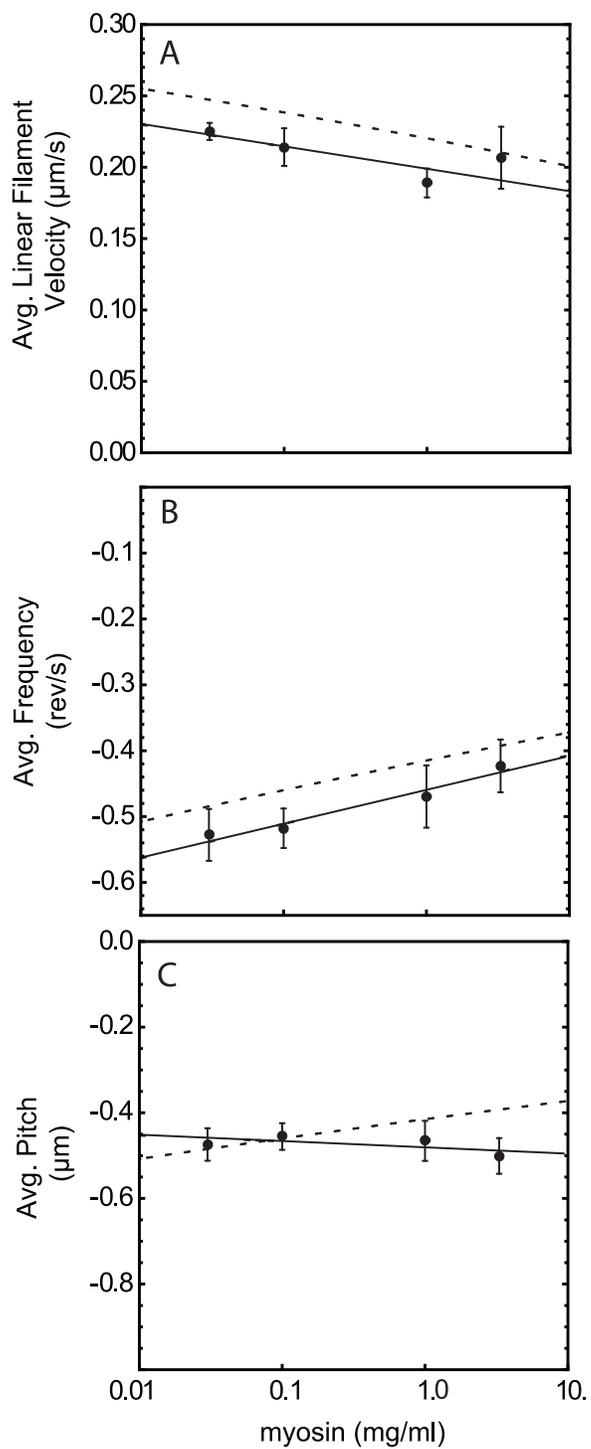

Figure 6:

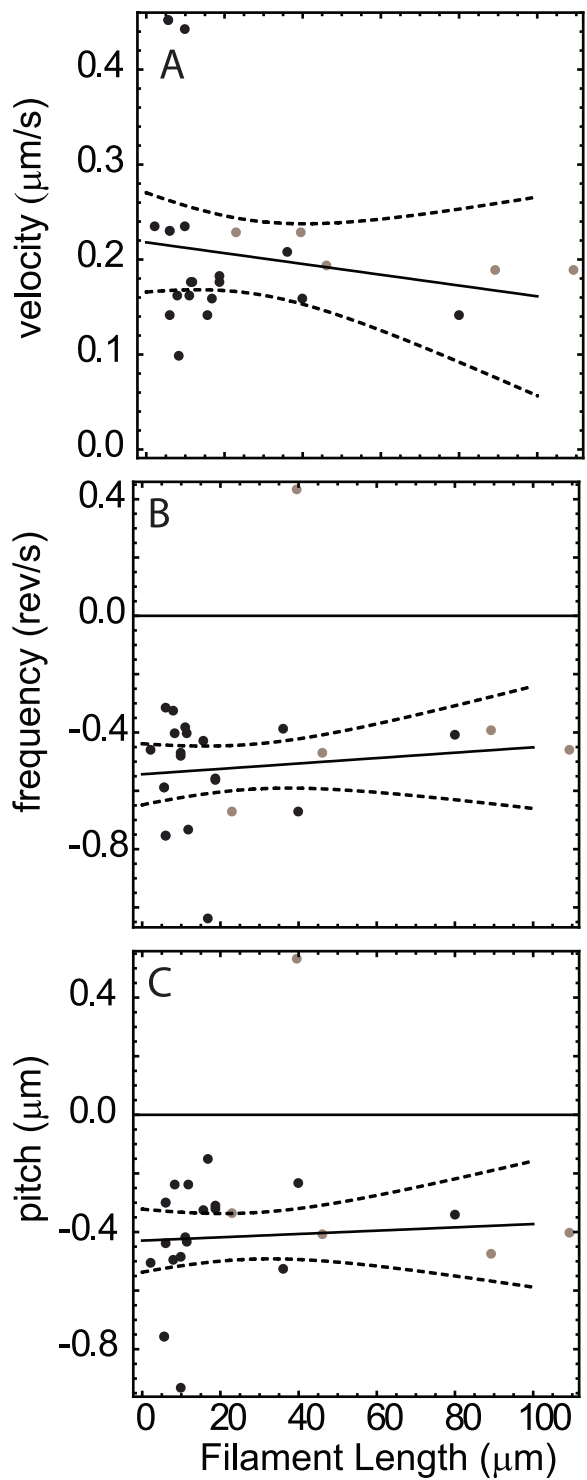

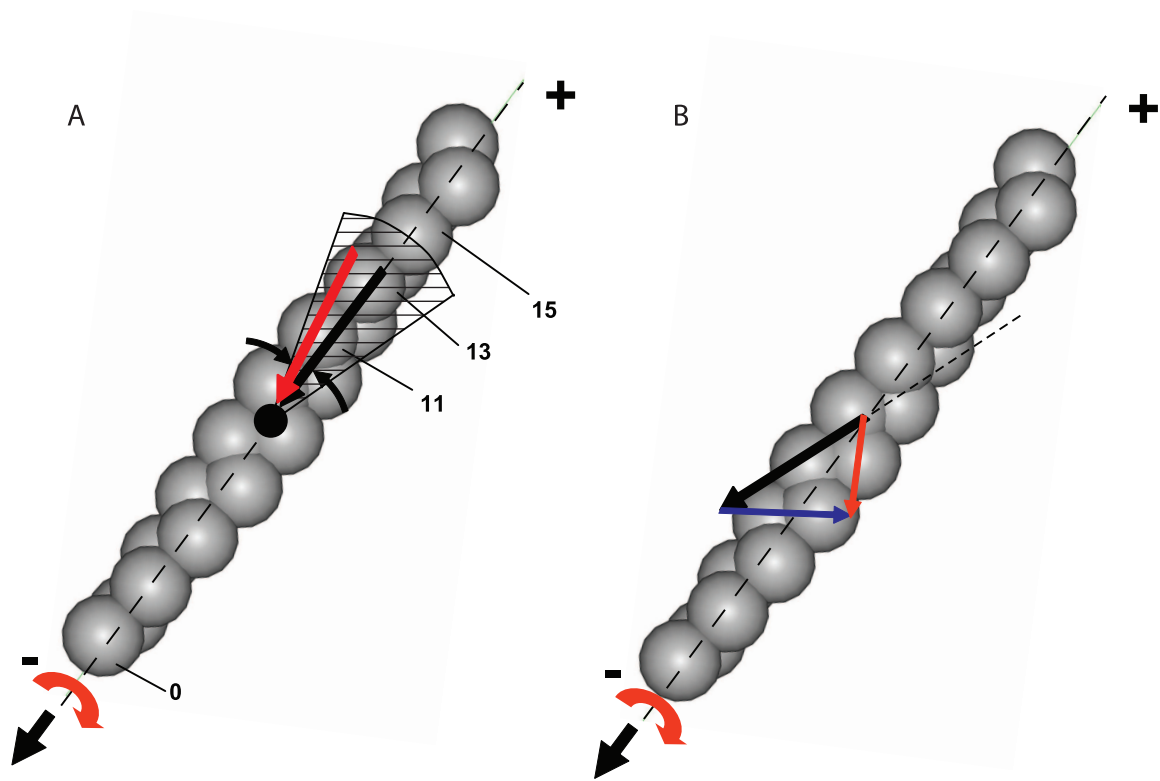

Figure 8: